\documentstyle[aps,prl,twocolumn,epsf,epsfig]{revtex}

\begin{document}
\draft
\twocolumn[\hsize\textwidth\columnwidth\hsize\csname @twocolumnfalse\endcsname

\title{Enhancing $T_c$ in ferromagnetic semiconductors}
\author{S. Das Sarma, E. H. Hwang, and D. J. Priour, Jr.}
\address{Condensed Matter Theory Center, Department of Physics,
University of Maryland, College Park, MD 20742-4111}

   
\maketitle

\begin{abstract}
We theoretically investigate disorder effects on the ferromagnetic
transition (``Curie'') temperature $T_c$ in dilute III$_{1-x}$Mn$_x$V
magnetic semiconductors (e.g., Ga$_{1-x}$Mn$_x$As) where a small
fraction ($x\approx 0.01-0.1$) of the cation atoms (e.g., Ga) are
randomly replaced by the magnetic dopants (e.g. Mn), leading to
long-range ferromagnetic ordering for $T<T_c$. We find that $T_c$ is a
complicated function of at least eight different
parameters including carrier density, magnetic dopant density, and
carrier mean free path; nominally macroscopically similar samples
could have substantially different Curie temperatures. We provide
simple physically appealing prescriptions for enhancing $T_c$ in
diluted magnetic semiconductors, and discuss the magnetic phase
diagram in the system parameter space.
\end{abstract}

\pacs{PACS numbers: 75.50.Pp,75.10.-b,75.10.Nr,75.30.Hx}
]

Diluted magnetic semiconductors (DMS), which are ferromagnetic for
$T<T_c$, are 
of great fundamental interest because they provide a unique
example of interplay among magnetism, disorder, carrier dynamics, and
transport properties. Starting with the early report of
carrier-induced global ferromagnetism first in In$_{1-x}$Mn$_x$As
\cite{InMnAs} and later \cite{GaMnAs} in Ga$_{1-x}$Mn$_x$As, with the
typical magnetic impurity concentration $x\approx 5\%$, there have been
many studies reporting ferromagnetism in DMS systems as disparate as
Ge$_{1-x}$Mn$_x$ \cite{GeMn}, Ga$_{1-x}$Mn$_x$N \cite{GaMnN},
Ga$_{1-x}$Mn$_x$P \cite{GaMnP}, TiO$_2$-Co\cite{TiOCo},
SnO$_2$-Co\cite{SnOCo}, and ZnCrTe \cite{ZnCrTe} among others. There is a
great deal of current research activity in the subject, and new DMS
materials with novel magnetic properties are likely to continue
appearing in the near future. In spite of this enormous 
activity, there is no current consensus  on the basic
magnetic model underlying DMS ferromagnetism. In fact, there is strong
disagreement in the literature about the nature of the carrier ground
state participating in the ferromagnetism --- in particular, whether
these carriers are free carriers in the parent-semiconductor bands
or are impurity (or defect)
band carriers created within the fundamental band gap of the
semiconductor by the magnetic dopant atoms is a matter of considerable
controversy in the literature \cite{SDS}. Experimentally the
situation turns out to be quite complex as the ferromagnetic
properties (in particular, the Curie 
temperature $T_c$) seem to depend very sensitively \cite{Anneal} on
the materials growth and processing (e.g. annealing) conditions, and
$T_c$ in nominally identical samples could differ substantially
depending on the precise details of sample preparation and
processing.

In this letter, we develop a theory for the prediction of $T_c$ in DMS
systems by focusing on the most-studied DMS system,
viz. Ga$_{1-x}$Mn$_x$As, as a function of various system parameters
(to be discussed below). Our work, which includes spatial disorder of
the system (by virtue of the random locations of the magnetic Mn
atoms), leads naturally to the conclusion that $T_c$
is a complicated (and in general unknown)
function of the system parameters; the number of independent
variables (at least eight as discussed below)
determining $T_c$, even within the minimal zeroth order effective
Hamiltonian approach, are sufficiently large that nominally
macroscopically identical samples may very well have significant 
variations in $T_c$, as is indeed observed. 
Some of the system parameters (e.g. the nature of spatial
disorder, various defects present in the real samples, the
impurity scattering potential, etc.) are generally unknown as a matter
of principle, and therefore a precise quantitative prediction for
$T_c$ as a function of the {\it known} system parameter 
(i.e. the Mn concentration $x$ and the carrier density $n_c$),
as is often done in the literature, could be quite meaningless
because
the same values of $x$ and $n_c$ may lead to
different $T_c$ values in different samples depending on the
details of various
defects/impurities/disorder in the system (which can be influenced,
for example, by sample annealing). 
Among the various possible defects,
As antisites 
and Mn interstitials 
are known to be
important. In addition, likely correlated clustering of Mn atoms
(instead of uncorrelated random positioning at Ga sites) is also
thought to be significant \cite{Timm02}. Experimentally, a strong
dependence of $T_c$ on the sample conductivity has been found with $T_c$
typically increasing with conductivity; samples with the 
highest (lowest) conductivity invariably have the highest (lowest)
$T_{c}$s, but whether this dependence arises entirely
from the higher conductivity samples having higher carrier densities
(as is commonly assumed in the literature) or longer mean free paths (MFP)
(or equivalently, higher mobility) or a combination of the two is
presently unknown. We therefore see that $T_c \equiv
T_c(n_i,n_c,l,J_0;N_{AS},N_I,N_C,J_{AF} \cdot \cdot \cdot)$ is a
function of at least eight different parameters 
due to the complexity of the problem.
The three obvious parameters, which have been discussed widely in the
literature, are the active magnetic moment density $n_i$ due to
substitutional Mn dopants, the carrier density $n_c$, and the carrier
MFP $l$ (as obtained from the $dc$-conductivity), and in
this paper we mostly focus on the dependence of $T_c$ on these three
parameters. The other relevant parameters, e.g. $N_{AS}$ (the As
antisite defect density), $N_I$ (the Mn interstitial defect density),
$N_c$ (a set of parameters defining the
clustering of Mn atoms or the correlation in their spatial positions),
and $J_{AF}$ (the direct short-range Mn-Mn antiferromagnetic exchange
interaction), are quantitatively important, but essentially unknown
(either experimentally or theoretically), and their explicit inclusion
in the theory for a quantitative comparison with the experimental
results is hence not particularly useful. 
However, we assume that the influence of
these parameters can be included qualitatively in the
theory by appropriately adjusting the parameters $n_i$, $n_c$, $l$,
and most importantly, the effective ferromagnetic coupling strength
$J_0$ which sets the overall energy scale
in the problem since we express $T_c$ in the units of $J_0$. We find
that even with just 
three independent parameters ($n_i$, $n_c$, $l$), the
problem is quite rich leading to many subtle possibilities. 

We use the standard RKKY-Zener effective magnetic model
\cite{AG} for describing the coupled carrier-local
moment system since we are interested in the so-called ``metallic''
DMS regime with itinerant carriers where $T_c$ is maximized. (The
insulating strongly localized DMS regime can also be ferromagnetic,
but the nature of the insulating DMS ferromagnetism, with typically
rather low $T_c$, is fundamentally different \cite{Kaminski} from the
metallic regime with higher $T_c$ being considered in this Letter.) We
note that while it is somewhat crude to characterize the metallic DMS
regime by a MFP $l$ extracted from the conductivity $\sigma
=n_ce^2\tau/m_c$ (and $l=v_F \tau$, where $\tau$ is the transport
relaxation time, $m_c$ the carrier mass, and $v_F$ the carrier Fermi
velocity), the extracted value of the MFP $l$ is rather
short (typically around one to a few lattice constants). Following our
earlier work \cite{Priour} we employ an effective magnetic model
Hamiltonian for interacting  
impurity moments where the carrier degrees of freedom have been integrated
out:
\begin{equation}
H = \sum_{i,j}J_F({\bf r}_{ij}) {\bf S}_i \cdot {\bf S}_j +
{\sum_{i,j}}'J_{AF}{\bf S}_i \cdot {\bf S}_j,
\label{Ham}
\end{equation}
where the subscripts F (AF) denote the ferromagnetic
(antiferromagnetic) part of the impurity spin ({\bf S}$_i$)
interaction and the prime in the second term implies that the
sum is restricted to nearest neighbors since the
short-range antiferromagnetic interaction only couples
nearest-neighbor Mn atoms (if there are any). 
We do not explicitly include the
direct antiferromagnetic exchange term in the results shown
in this paper assuming without any loss of generality (within our
theoretical scheme) that $J_{AF}$ effectively modifies 
(in fact, reduces) the carrier-mediated ferromagnetic
interaction. Since the 
interaction strength is the basic energy parameter in our effective
magnetic Hamiltonian, it makes little sense to keep two unknown
coupling parameters. We write Eq. (\ref{Ham}) with the direct
antiferromagnetic exchange term for our general qualitative discussion
of the DMS phase diagram as described below.

The first term in Eq. (\ref{Ham}), the carrier-mediated ferromagnetic
inter-impurity interaction, is of the RKKY form in our effective
model:
$J_F(r)= J_0[(2k_{F}r)\cos(2k_{F}r)-\sin(2k_{F}r)]/(k_Fr)^4$,
where $J_0$ is the fundamental ferromagnetic coupling parameter in the
problem (which implicitly includes all materials and band structure
information about the system) and  $k_F$ the carrier Fermi momentum 
(and $r = |{\bf r}_{ij}|$ the spatial separation between randomly
located substitutional Mn pairs in the GaAs lattice). 
We note one
important feature of $J_F(r)$: for high carrier density, $n_c
\ge n_i$, the oscillatory aspects of RKKY interaction come into play,
potentially suppressing DMS ferromagnetism --- 
the details of this suppression are an important topic of this work. A
straightforward Weiss mean-field treatment of the RKKY interaction,
neglecting all spatial disorder effects and assuming a continuum
virtual crystal approximation (VCA), was first carried out a long time ago
(and has recently been rediscovered in the DMS context) leading to: 
$T_c^{VCA} \propto n_in_c^{1/3}$.
This $T_c^{VCA}$ (with appropriate quantitative modifications arising
from band structure effects) has been extensively (and uncritically in
our view) used in the DMS literature for explaining and predicting
$T_c$ in DMS systems. In the current work we include in our 
calculation of $T_{c}$ spatial disorder
(thus relaxing the continuum VCA) and finite MFP, 
which are both significant in DMS materials.

\begin{figure}
\centerline{\epsfig{figure=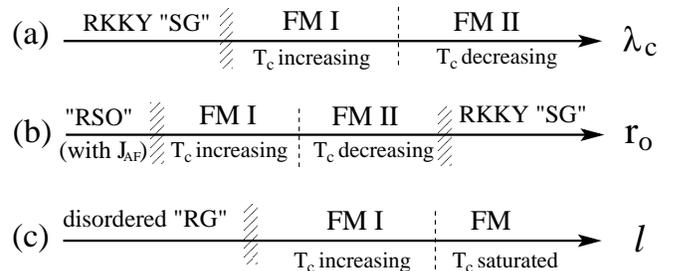,width=3.5in}}
\caption{DMS phase diagram. Note that the dashed lines between FM
  phases are {\it not} phase transition.
}
\label{Fig1}
\end{figure}

First we qualitatively discuss the DMS phase diagram as a function of
the variable $n_i$, $n_c$, and $l$, noting that the standard VCA
implies that the system is a ferromagnet for {\it
  all} values of $n_c$ and $n_i$ with $T_c^{VCA}$ increasing
monotonically with increasing 
impurity ($n_i$) and
carrier ($n_c$) density.  It is essential to include MFP
effects in the theory. The RKKY interaction has been calculated in the
presence of resistive scattering earlier in the literature
\cite{DeGennes}, and the modified RKKY interaction in
the presence of a finite MFP $l$ has the form 
$J_F(r;l) =  J_F(r)$ for $r \ll l$ and  
$J_F(r;l) =  J_0 \cos(\phi(r))/(2k_Fr)^3$ for $ r \gg l$,
where $J_{F}(r)$ is the standard RKKY formula and
$\phi(r)$ is a completely random function of $r$. 
The inclusion of short-range direct antiferromagnetic exchange
[Eq.~(\ref{Ham})], RKKY oscillations, and 
scattering/transport MFP effects  permits us 
to obtain the qualitative $T=0$ DMS phase diagram as
a function of three length variables $\lambda_c \sim n_c^{-1/3}$; $r_o
\sim n_i^{-1/3}$; $l$.  The schematic phase diagram depicted in 
Fig.~\ref{Fig1} 
assumes $\lambda_c$, $r_o$, and $l$ to be completely 
independent variables (which they cannot be in real systems). In each

\noindent
case shown in Fig.~\ref{Fig1} we assume the other variables to be 
{\it fixed} at some reasonable ``optimum'' values (which may not
always be experimentally possible).

In Fig.~\ref{Fig1}(a) the system is an ``RKKY spin-glass (SG)'' system at
very small values of $\lambda_c$ (equivalently very large values of
$n_c$). This arises from the frustration induced by the RKKY
oscillations which are dominant in the high $k_F$ ($n_c \gg n_i$) limit
(but are essentially absent in the dilute $n_c \ll n_i$ limit). As
$\lambda_c$ increases (i.e. $n_c \sim \lambda_c^{-3}$ decreases) the
``RKKY SG'' 
phase gives way to a ferromagnetic phase.  The Curie temperature of
this phase   
increases with decreasing density until the optimum $T_c$ is reached,
after which 
$T_c$ begins to decrease with decreasing carrier density.
In Fig.~\ref{Fig1}(b) we depict the DMS phase diagram
in magnetic impurity density $n_i$ ($\sim r_0^{-3}$) assuming
$l$ and $\lambda_c$ to be fixed at reasonable optimal values. For
small $r_0$ (large $n_i$) the nearest-neighbor direct
antiferromagnetic exchange between the Mn moments becomes important
and competes with the carrier-induced RKKY interaction,
leading to a random spin-ordered (RSO) 
non-ferromagnetic phase (which may or
may not be a SG phase). For larger $r_0$ we enter the
ferromagnetic phase with $T_c$ ($r_0$) similar to the
$T_{c}$ profile in Fig.~\ref{Fig1}(a); eventually we find a phase transition
to the RKKY SG phase for very large $r_0$ where $n_c \gg
n_i$. Thus, except for the antiferromagnetic
exchange induced ``RSO'' phase in the high impurity concentration
limit in Fig.~\ref{Fig1}(b), the phase diagrams of Fig.~\ref{Fig1}(a)
and (b) are essentially mirror images of each other as one shifts from
$n_c \gg n_i$ ($n_i \gg n_c$) to $n_c \ll n_i$ ($n_c \gg n_i$)
in Fig.~\ref{Fig1}(a) [(b)] respectively. Finally, in Fig.~\ref{Fig1}(c) we
discuss the phase diagram as a function of the MFP for
fixed $n_i$ and $n_c$ values. For very small $l$, the
carrier-induced ferromagnetic RKKY coupling is suppressed, and the
inter-impurity interaction has random sign, leading to a type of 
(non-RKKY) disordered ``random glassy'' (RG) 
non-ferromagnetic ground state which,
with increasing $l$, should make a phase transition to the
ferromagnetic phase. As one increases $l$ further the ferromagnetic
phase should initially be enhanced (i.e. $T_{c}$ rises with $l$) with
an eventual 
saturation of $T_c$ determined by the precise 
values of $n_c$ and
$n_i$. Thus it is readily evident 
that the `best' technique to
enhance $T_c$ would be to increase the MFP sitting at fixed
optimal 
values of $n_c$ and $n_i$. Although this is perfectly
reasonable as a matter of principle, it may be difficult to increase
the MFP without affecting $n_c$ and $n_i$.

Following this qualitative introduction to the DMS phase diagram on
the basis of the simple effective Hamiltonian
approach, we now consider the quantitative dependence of
$T_c(n_c,n_i,l)$ in DMS systems focusing on the well-studied
Ga$_{1-x}$Mn$_x$As system. To do this we carry out a thermal lattice
mean field treatment (treating exactly the spatial disorder of random Mn
locations at Ga substitutional sites in the zinc blende GaAs lattice)
of the effective Hamiltonian as described in \onlinecite{Priour}. Such a
treatment avoids the physically unrealistic 
assumptions 
\begin{figure}
\centerline{\epsfig{figure=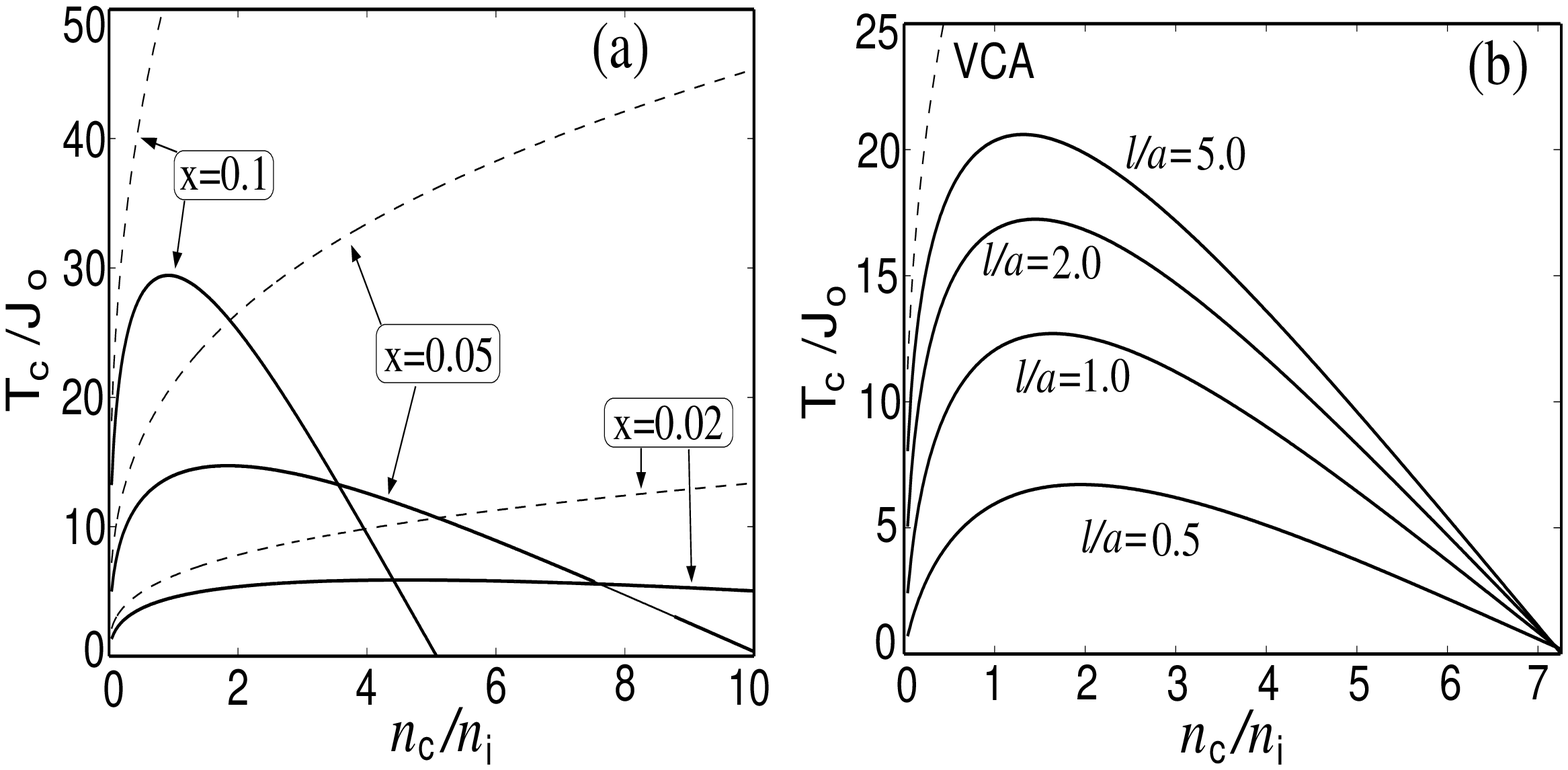,width=3.4in}}
\caption{Calculated $T_c$ (solid lines) as a function of $n_c/n_i$ (a)
  for different 
$x$ values and fixed MFP $l/a=5$, and (b) for different MFP $l/a$ and
fixed $x=0.07$. Dashed lines indicate the results from VCA.
}
\label{Fig2}
\end{figure}
\noindent
of the 
continuum VCA involved in deriving $T_c$,
and should be an excellent approximation for
obtaining $T_c$ because of the very large coordination number in the
fcc zinc blende GaAs lattice structure. As emphasized by us
elsewhere, this lattice mean field theory (LMFT)
can essentially be carried
out for infinite size systems, thus avoiding the finite size
complications inherent in direct Monte Carlo simulations which are
{\it not} particularly well-suited for rapidly determining $T_c$.

In Fig.~\ref{Fig2} we present our theoretical results for
$T_c(n_c,n_i,l)$. As noted earlier, we have ignored the direct 
short-range antiferromagnetic exchange interaction in obtaining 
these results.
We also ignore interstitial defects (and antisite As) in
this calculation; we assume that $n_i$ and $n_c$ are the {\it effective}
active local moment and hole density respectively, which already
incorporate various defect effects.

In Fig.~\ref{Fig2}(a) we show the calculated $T_c$ as a function of
$n_c/n_i$ for several values of $x$ in a ``highly'' metallic system
($l/a =5$, where $a$ is the GaAs lattice constant), whereas in
Fig.~\ref{Fig2}(b) we show $T_c(n_c/n_i)$ for a fixed value of $x$
($=0.07$) but for several different values of $l/a$. We show the
simple continuum VCA result
in each case for the sake of comparison. It is
obvious that the simple theory of $T_c^{VCA}$ is qualitatively
incorrect for large $n_c/n_i$ 
where  $T_c$ actually reaches a maximum
and then decreases with
increasing carrier density (due to the frustration inherent in the
RKKY oscillations playing a role for $n_c/n_i >1$ or
equivalently $k_Fr_0 >1$) in contrast to the erroneous claim (made
extensively in the literature) that $T_c(n_c)\propto n_c^{1/3}$ (as
obtained from continuum VCA) would
continue increasing monotonically with carrier density. It is,
however, important to note that our results presented in
Fig.~\ref{Fig2} indicate a fairly large regime of (qualitative and
even semi-quantitative) validity of the simple VCA
with the appropriate numerical
modification of $J_0$ which is an adjustable parameter merely setting
the scale of energy in our theory. In particular, the simple VCA theory
remains valid up to $n_c/n_i \sim 0.5$, and perhaps even up to $n_c/n_i
\sim 1.0$ depending on the Mn content (i.e. $x$). 
The optimum $n_c/n_i$ value
where $T_c$ is maximum decreases 

\begin{figure}
\centerline{\epsfig{figure=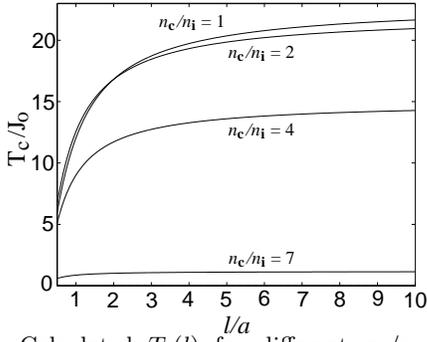,width=2.2in}}
\caption{Calculated $T_c(l)$ for different $n_c/n_i$ and fixed $x=0.07$.
Note $T_c^{VCA}/J_0=$32.6, 41.6, 52.2, 63.1 for $n_c/n_i=$1, 2, 4, 7,
respectively.}
\label{Fig3}
\end{figure}
\noindent
as Mn concentration increases, and
this indicates that, for a given value of the effective coupling
$J_0$, $T_c$ cannot really be arbitrarily increased just by increasing
the carrier density (although increasing carrier density by co-doping
with non-magnetic impurities\cite{Park} should enhance
$T_c$ somewhat), but typically $T_c$ is optimum for $n_c/n_i \sim
1$. This suggests that the current popular wisdom of trying to
enhance $T_c$ for GaMnAs (and other DMS materials) simply by
increasing carrier density would not work much beyond $T_c \sim 300$K
(since the current maximum $T_c$ is around 170K).

Our results for $T_c(l)$ presented in Fig.~\ref{Fig3} indicate one
possible strategy for enhancing $T_c$. As can be seen in Fig.
\ref{Fig3}, $T_c$ increases monotonically with increasing MFP
eventually saturating at the maximum possible $T_c$ for a given
value of $n_c/n_i$ (which is somewhat below the corresponding
$T_c^{VCA}$). Thus a clear strategy to enhance $T_c$ is to
optimize  $n_c/n_i$  for a given Mn content such that one is
at or near the optimum carrier density (i.e. near the maximum in
Fig.~\ref{Fig2}), $n_c/n_i \sim 0.5-2$ depending on $x$, and then to
enhance the carrier MFP by reducing scattering effects
through a systematic improvement of sample quality. Experimentally, it
is now established \cite{Conductivity} that enhancing conductivity by
improving sample quality (e.g. via annealing) can substantially
increase $T_c$, but
the conductivity $\sigma$ 
depends both on the 
carrier density $n_c$ and the
MFP $l$, and it has almost universally been assumed that the increase
of $T_c$ due to enhanced $\sigma$ arises entirely from the increasing
carrier density, whereas we find that $T_c$ improvement arises both
from increasing $n_c$ and $l$.

\begin{figure}
\centerline{\epsfig{figure=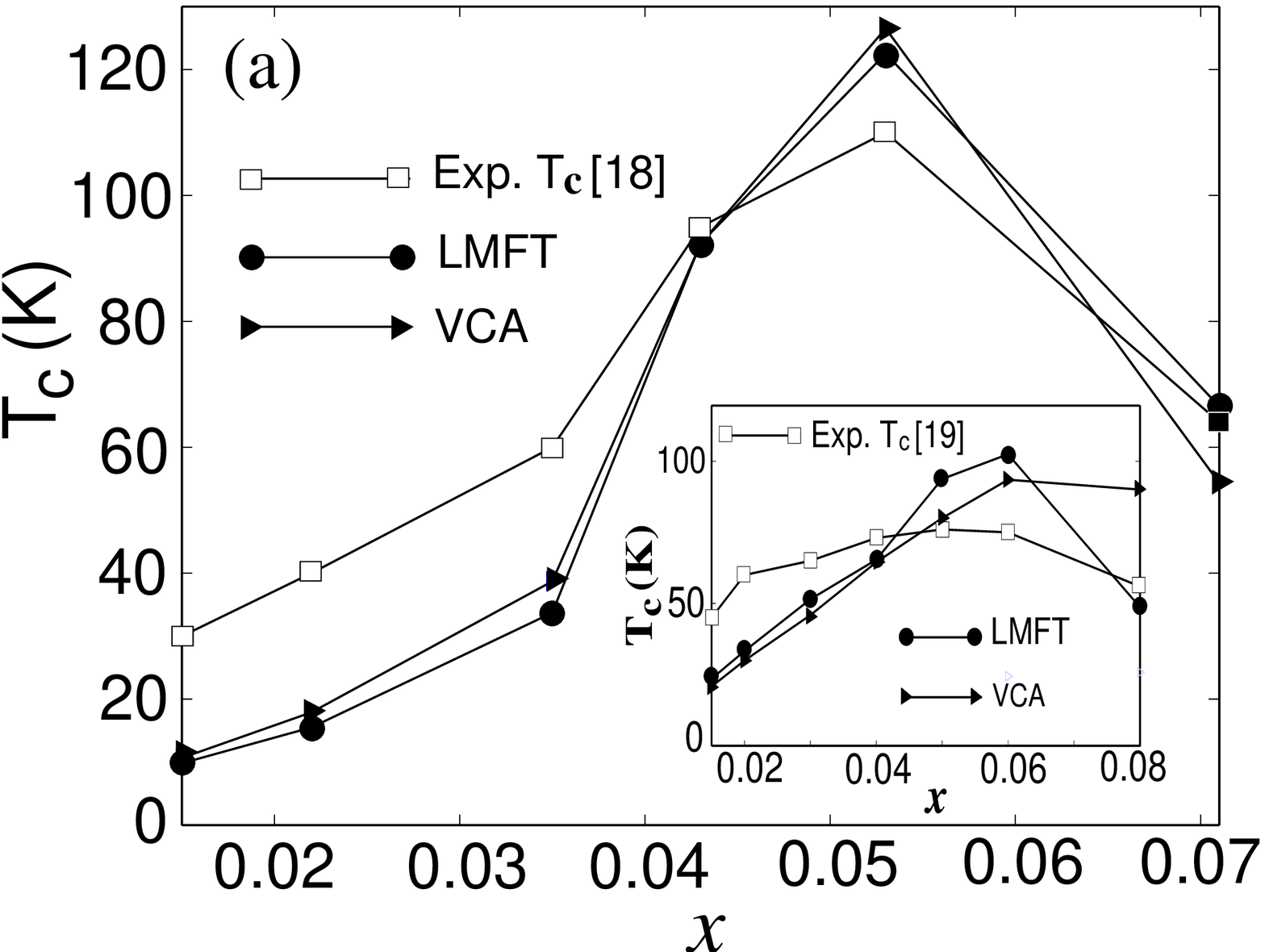,width=2.5in}}
\centerline{\epsfig{figure=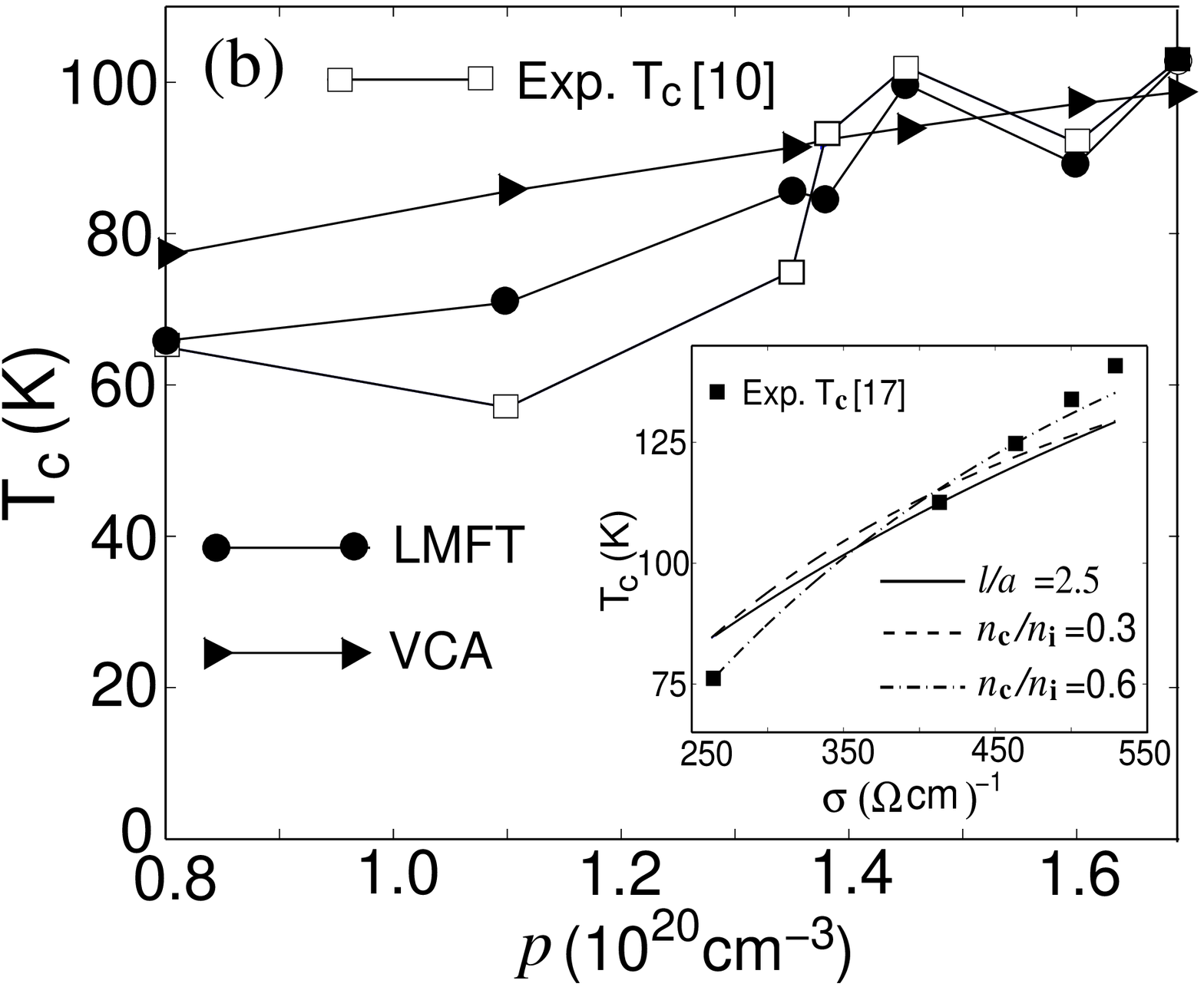,width=2.5in}}
\caption{(a) Comparison of $T_c$ with experimental results of Ref. [18] and
  [19] (inset) as a function of $x$. (b) Comparison of $T_c$ with
  experimental results of Ref. [10] as a function of hole density and
  those of Ref. [17] as a function of conductivity (inset).
}
\label{Fig4}
\end{figure}

Finally, in Fig.~\ref{Fig4} we show a direct comparison between our
theory and recent GaMnAs experimental results from several different
groups \cite{Anneal,Conductivity,Matsukura,Edmonds}. For each set of
results in Fig.~\ref{Fig4} we have extracted 
$T_c(n_c,n_i,\sigma(l))$ from the relevant experimental work as
described in the figure captions. It is obvious that the
experimental 
results are well-described by the theory; the agreement can be made
essentially exact by slightly adjusting $n_i$ and/or by choosing
slightly different $J_0$ for different values of $x$, both of which
may be reasonable since Mn interstitials

\noindent
(whose density may very well
be a function of $x$) are likely to affect the value of effective
coupling by modifying the number of active Mn moments participating in
global ferromagnetism and by introducing some direct Mn-Mn
antiferromagnetic exchange coupling. It is also evident from
Fig.~\ref{Fig4} that the continuum VCA, while qualitatively
reasonable in some regimes of the parameter space, does not provide a
good quantitative description for the experimental results.

In conclusion, we have discussed the DMS phase diagram,  
taking into account effects of carrier-mediated
ferromagnetic and antiferromagnetic coupling between the impurity
moments as well the frustration arising from the  RKKY
oscillations, finding that ferromagnetism is only one of four distinct
magnetic phases possible in the disordered system. We have also carried
out a detailed theoretical $T_c$ calculation as a function of magnetic
impurity concentration, carrier density, and conductivity (i.e. MFP)
including full effects of spatial disorder and randomness, finding
that the maximum $T_c$ is obtained for an optimum carrier density
$n_c/n_i \sim 0.5-2$ depending on the Mn concentration. We have shown
that for $T_c$ to be enhanced at given values of $n_c$ and $n_i$, one
needs to increase the MFP as much as possible. 

This work is supported by the US-ONR and DARPA.

\end{document}